\documentclass[9pt,twocolumn,twoside]{pnas-new}

\setboolean{displaywatermark}{false}

\templatetype{pnasresearcharticle} 

\usepackage{balance}
\usepackage{mathrsfs}
\usepackage{amsmath}
\usepackage{amsfonts}
\usepackage{mathtools}
\usepackage{tikz}
\usepackage{graphicx}
\usetikzlibrary{positioning}
\usepackage[export]{adjustbox}
\usepackage{rotate}
\usepackage{dsfont}
\usepackage{relsize}
\usepackage{epstopdf}
\DeclareGraphicsExtensions{.jpg,.png,.pdf,.eps}
\usepackage{pstricks}
\usepackage{multirow}
\usepackage{color}
\usepackage{import}
 
\usepackage{dsfont}
\usepackage{relsize}

\usepackage{filecontents}
\usepackage{multibib}
 
\usepackage{epstopdf}

\newcolumntype{Y}{>{\centering\arraybackslash}X}
\newcolumntype{A}{>{\hsize=.1\hsize}Y}
\newcolumntype{B}{>{\hsize=.2\hsize}Y}
\newcolumntype{C}{>{\hsize=.4\hsize}Y}
\newcolumntype{D}{>{\hsize=.2\hsize}Y}

\newcommand{\rb}{\mathbf{r}}

\newcommand{\vb}{\mathbf{v}}

\newcommand{\xb}{\mathbf{x}}

\makeatletter
\def\@fnsymbol#1{
	\ensuremath{\ifcase#1\or
	*\or				
	\ddagger\or			
	\mathsection\or		
	\dagger\or			
	\mathparagraph\or	
	\|\or				
	**\or				
	\ddagger\ddagger\or	
	\dagger\dagger		
	\else\@ctrerr\fi}}
\makeatother

\title{Motility-induced buckling and glassy dynamics regulate three-dimensional transitions of bacterial monolayers}
 
\author[a,1]{Sho C. Takatori}
\author[b,c,1]{Kranthi K. Mandadapu}
\affil[a]{Department of Chemical Engineering, University of California, Santa Barbara, Santa Barbara, CA 93106}
\affil[b]{Department of Chemical and Biomolecular Engineering, University of California, Berkeley, Berkeley, CA 94720}
\affil[c]{Chemical Sciences Division, Lawrence Berkeley National Laboratory, Berkeley, CA 94720}

\leadauthor{Takatori} 

\correspondingauthor{\textsuperscript{1} To whom correspondence should be addressed: stakatori@ucsb.edu, kranthi@berkeley.edu}


\begin{abstract}
Many mature bacterial colonies and biofilms are complex three-dimensional (3D) structures.  
One key step in their developmental program is a transition from a two-dimensional (2D) monolayer into a 3D architecture.  
Despite the importance of controlling the growth of microbial colonies and biofilms in a variety of medical and industrial settings, the underlying physical mechanisms behind single-cell dynamics, collective behaviors of densely-packed cells, and 3D complex colony expansion remain largely unknown.  
In this work, we explore the mechanisms behind the 2D-to-3D transition of motile \textit{Pseudomonas aeruginosa} colonies; we provide a new motility-induced, rate-dependent buckling mechanism for their out-of-plane growth. 
We find that swarming of motile bacterial colonies generate sustained in-plane flows. 
We show that the viscous shear stresses and dynamic pressures arising from these flows allow cells to overcome cell-substrate adhesion, leading to buckling of bacterial monolayers and growth into the third dimension.  
Modeling bacterial monolayers as 2D fluid films, we identify universal relationships that elucidate the competition between in-plane viscous stresses, pressure and cell-substrate adhesion.  
Furthermore, we show that bacterial monolayers can exhibit crossover from swarming to kinetically-arrested, glassy-like states above an onset density, resulting in distinct 2D-to-3D transition mechanisms.  
Combining experimental observations of \textit{P. aeruginosa} colonies at single-cell resolution, molecular dynamics simulations of active systems, and theories of glassy dynamics and 2D fluid films, we develop a dynamical state diagram that predicts the state of the colony, and the mechanisms governing their 2D-to-3D transitions.  
\end{abstract}
    
\begin{document}

\maketitle

\ifthenelse{\boolean{shortarticle}}{\ifthenelse{\boolean{singlecolumn}}{\abscontentformatted}{\abscontent}}{}

%
%

\section*{Introduction} \label{sec:sec_intro}
We address the rate-dependent, three-dimensional (3D) transitions in motile confluent bacterial monolayers using \textit{Pseudomonas aeruginosa} as our model organism. \textit{P. aeruginosa} is a gram-negative, rod-shaped, motile bacterium that can cause serious infections in humans with existing diseases or conditions, including cystic fibrosis and traumatic burns \cite{Lyczak00,Gellatly13,Faure18,Lopez15,Bjarnsholt13}.  
Fiber-like motorized appendages called type-IV pili enable twitching motility and lead to complex swarming patterns on soft substrates \cite{Kearns10,Rashid00,Gloag13,Verstraeten08}.  \smallskip

\begin{figure*}[b!]
\centering
\includegraphics[scale=0.58]{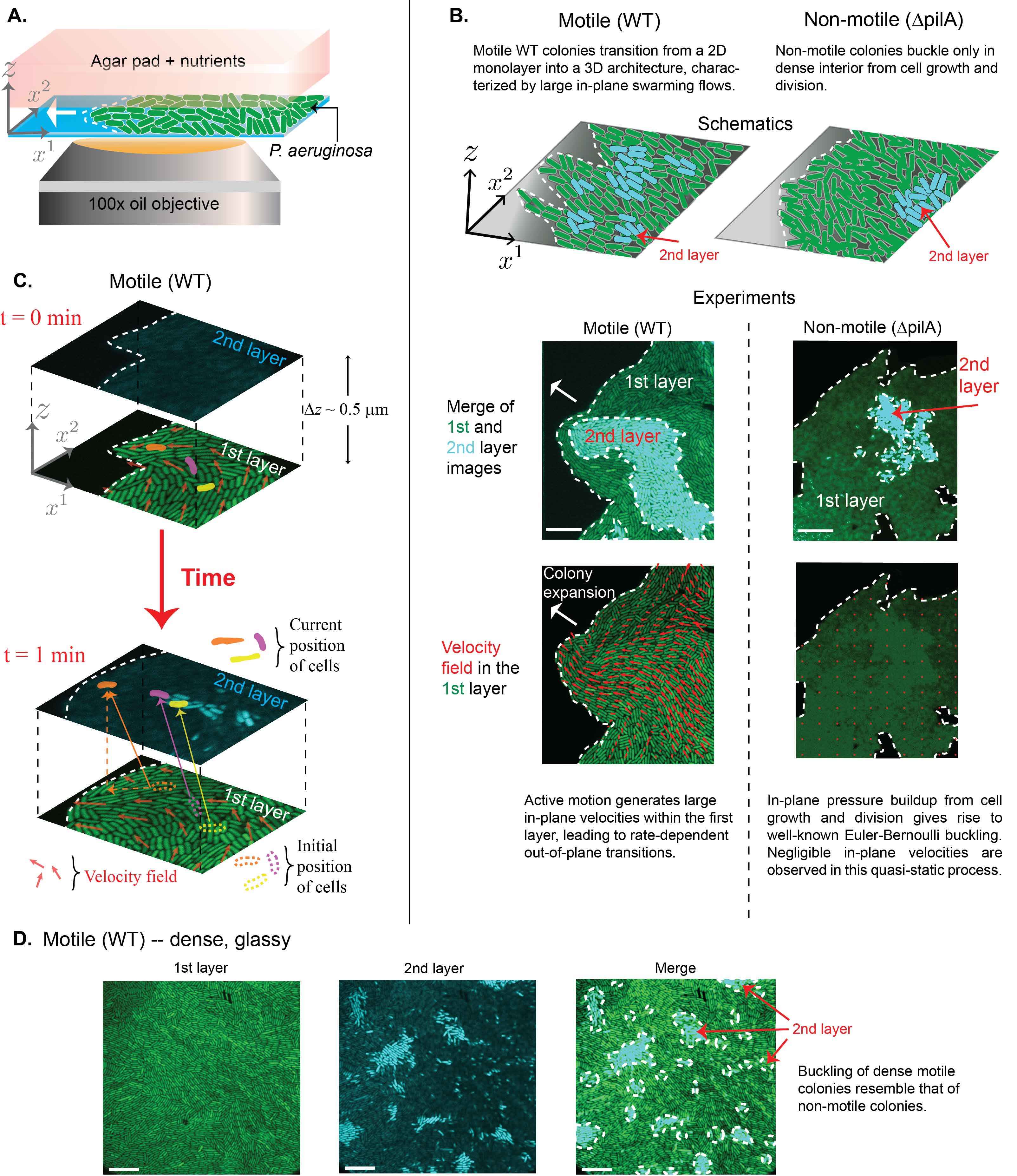}
\caption{Motile colonies of \textit{P. aeruginosa} PAO1 transition from a two-dimensional (2D) monolayer into a three-dimensional (3D) architecture.  
(A) Schematic of under-agar experimental assay for observing confluent bacterial monolayers using confocal microscopy.  Time-lapse, z-stack images are taken to observe both first and second bacterial layers.  (B) Out-of-plane transitions of individual bacteria from 2D monolayers (cyan) is observed for motile (wild type WT, left column) and non-motile ($\Delta$pilA, right column) strains. For WT, these transitions occur in regions of large in-plane collective swarming flows, especially near the colony's leading edges.  In contrast, non-motile strains do not exhibit any flows within the colony, and transitions occur within the interior of the colony where cell density increases slowly via cell growth and division.  Snapshots of the experiments show a z-stack superposition of the first layer (in green) and a second layer (pseudo-colored in cyan).  Velocity fields within the monolayer (below) are shown with red arrows.  (C) Nucleation events corresponding to 2D to 3D transitions in motile colonies are shown by tracking three bacteria over the course of a minute.  Individual bacteria translate in the monolayer by several microns and collide into multiple neighbors before transitioning out of plane.  Multiple bacteria appear in the second layer within a minute.  (D) Snapshots of the first layer (in green), second layer (pseudo-colored in cyan), and merged images of WT colony, at large cell density.  Large densities suppress active swarming and the colony enters into a kinetically arrested glassy state.  Individual cells can be observed to transition into the second layer from growth and division, away from the leading edges and in the interior (see SI Movie).  Thus, dense motile colonies and associated 2D-to-3D transitions resemble that of non-motile mutant colonies. All scale bars are $10 \mu$m.}
\label{fig:expts-1}
\end{figure*}

By performing an under-agar assay with \textit{P. aeruginosa} strain PAO1 (see Fig.~\ref{fig:expts-1}A and Methods), we follow the transition of cells from a 2D planar monolayer into 3D at single-cell resolution using confocal microscopy. 
Careful inspection of wild-type (WT) colonies revealed collective twitching motion in the swarming state, marked by characteristic in-plane flows (Fig.~\ref{fig:expts-1}B), including linear flows like simple shear flow and nonlinear flows such as squeeze flow (see supplementary information (SI) movie S1). 
These flows were especially pronounced at the migrating front of the expanding colony.  
Interestingly, we observed out-of-plane transitions of individual cells from a 2D monolayer into the third dimension, and that these nucleation events coincided with the collective flows within the colony (see SI movies S1-S3 and Fig.~\ref{fig:expts-1}B).  
Tracking the trajectories of several bacteria leading to a nucleation of a second layer (as shown in Fig.~\ref{fig:expts-1}C) indicates that these active in-plane flows may be coupled to its out-of-plane transitions.  
These in-plane flows and nucleation events occur on time scales much smaller than cell growth and division. \smallskip

In contrast, a non-motile mutant (pilA mutant) colony migrates slowly with no collective in-plane flows, and does not exhibit any out-of-plane transitions near the migrating edges of the colony (see SI movies S4-S5).  
Instead, we observe transitions into the third dimension only in the colony interiors (Fig.~\ref{fig:expts-1}B).  
Furthermore, experiments of WT cells packed at very large cell density in Fig.~\ref{fig:expts-1}D show that dense motile colonies can enter into a jammed, glassy-like state leading to kinetic arrest (see SI movies S6-S7). 
The 2D-to-3D transitions of these dense WT colonies resemble that of the non-motile strain with sudden nucleation events leading to out-of-plane growth again in the interior (see Fig.~\ref{fig:expts-1}D). 
Nucleation of non-motile bacteria \cite{Yan19,Beroz18,You19,Dell18,Grant14,Boyer11,Farrell13} and confluent mammalian tissues \cite{Saw17,Shraiman05} into the third dimension arise from a known phenomenon of buckling due to cell division and growth-induced stresses, and occurs on timescales much larger than the motility-induced transition events we observe in the WT strain in the swarming state, whose mechanisms remain yet unknown. \smallskip

Taken together, these observations point to an interplay between cell density, active motility, and 3D growth of motile bacterial monolayers. 
Although it is now clear that such an interplay should exist, there are many unanswered fundamental questions: 
(i) What are the mechanisms involved in out-of-plane growth of a motile swarming colony; 
(ii) How does cell density and motility affect the state of the colony (i.e., swarming vs. glassy); and (iii) How does the state of the colony govern the 2D-to-3D transitions?  
By combining experiments, molecular dynamics simulations, and theories of glassy dynamics and fluid films, we set out to fundamentally understand the state of the bacterial colonies and provide novel mechanisms of out-of-plane colony growth. 

%
%
\section*{Swarming vs. Glassy colonies}
We begin by addressing the physical state of the bacterial colony at a collective level as a function of cell density.
As observed experimentally in Fig.~\ref{fig:expts-1}D, upon increasing cell density, we find that the motile colonies undergo a crossover from a state of swarming to a state marked by a dramatic slow down in their dynamics specifically in their cellular displacements.  
In particular, colonies prepared at high densities resemble that of disordered glassy states in supercompressed colloidal or supercooled molecular systems \cite{Chandler10,Keys11,Hunter12,weeks2000three,Song08,berthier2011dynamical} (see SI movie S8). 
It is well-known that glassy colloidal systems at high densities, or molecular systems at low temperatures, exhibit dynamic heterogeneity \cite{berthier2011trend, weeks2000three, Keys11, kob1997dynamical, dauchot2005dynamical}, i.e., intermittent motions in particle displacements and spatial correlations in the mobility of individual particles when observed over a small period of time.
This is in contrast to a low-density colloidal fluid, or a high-temperature molecular liquid, where the dynamics is homogeneous and uniform.\smallskip

Figure~\ref{fig:expts-glass}A shows experimental trajectories of the bacterial monolayer at sufficiently low and high density, color coded by displacements of particles based on body length. 
At low density, it can be seen that dynamics in the monolayer is spatially homogeneous where most of the bacteria displace by a body length over a small period of time $\Delta t$. However, at high density there exist spatially distinct mobile and immobile regions, thus exhibiting dynamical heterogeneity \cite{berthier2011trend, weeks2000three, Keys11, dauchot2005dynamical}.
Emergence of dynamic heterogeneity can also be seen in the displacement of individual bacteria $|{\mathbf{r}_i(t) -\mathbf{r}_i(0)}|$ in Fig.~\ref{fig:expts-glass}B. 
Here, one can see continuous motion of bacteria at sufficiently low densities resembling that of a liquid, while bacteria at high densities exhibit hopping events similar to that of a colloidal or molecular glass, thus making the bacterial monolayer an active, glassy colloidal system. 
The time scale $\Delta t$, referred to as an ``instanton time'', is defined as the time taken by a particle to make significant displacement from its current location \cite{Keys11}, as shown in Fig.~\ref{fig:expts-glass}B.
This kind of slowdown leading to dynamical arrest and dynamic heterogeneity upon increasing density has also been observed in other living systems such as confluent mammalian monolayers \cite{Angelini11,Garrahan11,Puliafito12}.\smallskip

When a glass-forming molecular liquid is cooled below an ``onset'' temperature, its relaxation time, defined as the time decay of the density fluctuations, changes from being Arrhenius to super-Arrhenius with regards to temperature \cite{angell1995formation, elmatad2009corresponding, ediger1996supercooled, debenedetti2001supercooled, Keys11, garrahan2003coarse}.
Analogously, when colloidal systems are compressed above an ``onset'' pressure $\Pi_0$, their relaxation also exhibits a crossover from being ``Arrhenius'', i.e., exponentially dependent on pressure, to being ``super-Arrhenius'' in pressure \cite{Isobe16}. There exist different perspectives to describe glassy dynamics which explain the significant slow down in the relaxation times as a function of temperature or pressure \cite{berthier2011theoretical}.
In this work, we use the perspective of dynamical facilitation (DF) theory \cite{Chandler10,garrahan2003coarse,Keys11} to characterize the active glassy dynamics of motile bacterial colonies, and the crossover from swarming to glassy state. 
DF theory has been successful in predicting the relaxation behaviors of systems including molecular liquids \cite{elmatad2009corresponding}, mixtures of glass forming super-cooled liquids \cite{katira2019theory}, atomistic systems containing binary mixtures of particles \cite{Keys11}, and binary hard-disc systems in two-dimensions \cite{Isobe16}. 
DF theory fundamentally takes into account the dynamical heterogeneity in terms of spatial heterogeneities, also referred to as ``excitations'' or ``soft-spots'', which represent mobile regions observed in an instanton time $\Delta t$.  
These excitations then facilitate the motion of nearby regions in a hierarchical manner, leading to a super-Arrhenius ``parabolic'' form for the relaxation time $\tau$ as a function of pressure $\Pi$,  i.e., $\tau(\Pi) \sim \exp([\kappa(\Pi - \Pi_0)]^2)$, where $\kappa$ is a system dependent energy scale, and $\Pi_0$ is the onset pressure \cite{Isobe16, Keys11}. See SI Chap.~II.1 for a detailed description of the DF theory for passive systems. \smallskip

To characterize the dynamic heterogeneity and to extend the applicability of DF theory for bacterial monolayers, we created a molecular model containing bi-disperse hard disks based on the ``active'' Brownian particle model, a paradigmatic model commonly studied in the emergent field of active matter \cite{Tailleur08,Marchetti13,Cates13,Ramaswamy10,Lauga09}.
Briefly, the model considers a suspension of self-propelled disks of diameter $\sigma$ that translate with a swimming velocity $U_0$, and their direction of motion is subject to rotational diffusion characterized by a reorientation time $\tau_\text{R}$.  We defined a nondimensional P\'eclet number as $Pe \equiv U_0 \tau_\text{R} / \sigma$, which is a ratio of the run length of the particle over its size.
Unlike experiments on real colonies, molecular simulations enable careful control over packing fraction $\phi$ and activity parameter $Pe$, and thus are well-suited to test and extend the applicability of DF theory to active colloidal systems and therefore to bacterial colonies. 
Figure~\ref{fig:expts-glass}C shows dynamical heterogeneity in our active molecular models at high area fractions $\phi$, that exemplify the behaviors of motile bacterial colonies at high densities (shown in Fig.~\ref{fig:expts-glass}A; also see SI movie S9). 
This shows that our active molecular models are reasonable representations of real bacterial colonies in their glassy states.
Extending the DF theory to the case of active systems yields the relaxation time of the system to be 
\begin{align}
    \ln\Big(\frac{\tau}{\tau_0}\Big) & = E_\text{a}(\Pi - \Pi_0), \hspace{63pt} \text{ if } \Pi \leq \Pi_0 \label{eq:modi-para-1} \\
    &= [\kappa(\Pi - \Pi_0)]^2 + E_\text{a}(\Pi - \Pi_0), \text{ if } \Pi > \Pi_0, \label{eq:modi-para-2}
\end{align}
where $\tau_0$ is the relaxation time at the onset pressure, and the quantities $E_{\text{a}}, \Pi_0$, and  $\kappa$ depend on the activity $Pe$; see SI Chap.~II.2.1 for a detailed analysis of the extension of the DF theory to active glassy systems. \smallskip
 
To verify the predictions from Eqs.~(\ref{eq:modi-para-1})-(\ref{eq:modi-para-2}), we estimate the relaxation times in the molecular model for various activities by studying the time decay of the density correlations described by the dynamic strucure factor 
\begin{equation}\label{eq:dyn-str-fac}
    F_s(\mathbf{k},t) = \frac{1}{N}\Big\langle \sum_i \exp(i \mathbf{k} \cdot[\mathbf{r}_i(t) - \mathbf{r}_i(0)] ) \Big\rangle.  
\end{equation}
Here, $\mathbf{k}$ is the wave-vector corresponding to the peak of the static structure factor, and $\langle \cdot \rangle$ denotes an ensemble average. 
Figure~\ref{fig:DF theory}A shows $F_s(\mathbf{k},t)$ for various fractions $\phi$, and therefore various pressures $\Pi$, for a given activity $Pe$. 
It shows that the long time decay of the correlation functions changes from being exponential at low fractions to stretched exponential at high fractions, resembling the dynamical correlation functions in standard glassy systems \cite{debenedetti2001supercooled, kob1999computer, horbach1998molecular}. 
Furthermore, defining the relaxation time $\tau$ as the time scale for the decay of $F_s(\mathbf{k},t)$ to a value of $0.1$, we find a remarkable crossover from an ``Arrhenius'' dependence on $\Pi$ at low pressures to a ``super-Arrhenius'' dependence above the onset pressure (see Fig.~\ref{fig:DF theory}B). 
Figure~\ref{fig:DF theory}B also shows that the parabolic form in \eqref{eq:modi-para-1} is in excellent agreement with the relaxation times obtained from $F_s(\mathbf{k},t)$ above the onset pressure $\Pi_0(Pe)$ for various $Pe$, thus demonstrating the universal features of glassy dynamics at all activities. 
Note that $\kappa$, characterizing the energy scale associated with glassy dynamics, in the parabolic form is not a fit to the data, but rather predicted from microscopic calculations of energy barriers associated with excitations (see SI Chap.~II.2.1). \smallskip
 
Molecular models of active Brownian disks provide an excellent system for extension of the theory of glassy dynamics to active colloidal systems. 
However, the active liquid state in these systems below the onset density does not result in a swarming state consisting of flocks as observed in experiments of bacterial colonies, though there exist significant similarities in the glassy state.
This is due to the lack of alignment interactions, which are required to stabilize flocking behaviors \cite{vicsek1995novel, gregoire2004onset, mahault2018self, epstein2019statistical}. 
To this end, we extend our studies to systems consisting of active 2D polydisperse spherocylinders, with polydispersity representative of \textit{P. aeruginosa} colonies (see SI Chap.~II.3.2). 
In this case, there exist alignment interactions between individual spherocylinders and we find that the system exists in a swarming state below the onset density, and again exhibits a crossover into a glassy state above the onset density (see SI Fig.~2.15).
We further extend the DF theory to polydisperse spherocylinder systems, and show that Eqs.~(\ref{eq:modi-para-1}) and (\ref{eq:modi-para-2}) accurately represent the relaxation behaviors. \smallskip

Motivated by the success of the extended DF theory to active Brownian disks and polydisperse spherocylinder systems, we now study its applicability to $\textit{P. aeruginosa}$ colonies. 
To this end, we calculate the dynamic structure factor at multiple densities using \eqref{eq:dyn-str-fac} by tracking the trajectories of individual bacteria in the colony using confocal microscopy (see Methods).
Relaxation times collected from \textit{P. aeruginosa} colonies at different densities again exhibit a similar crossover from swarming to glassy states, as shown in Fig.~\ref{fig:DF theory}C. 
Furthermore, relaxation time behaviors of the colonies above the onset density can be quantitatively explained using the parabolic form \eqref{eq:modi-para-2} (SI Chap.~II.4). 
Taken together, Figs.~\ref{fig:DF theory}B and \ref{fig:DF theory}C exhibit a dynamical state diagram that determines the state of the colony (swarming vs. glassy) as a function of motility strength and cell density.
If cell densities are less than the onset density, and therefore less than the corresponding onset pressure $\Pi_0(Pe)$, the system exists in a swarming state, where transitions into 3D appear to be correlated with active in-plane flows (Figs.~\ref{fig:expts-1}B and \ref{fig:expts-1}C). 
If above the onset pressure, the bacterial monolayer enters into an arrested super-compressed or glassy state, and the out-of-plane mechanism resembles the colonies of a non-motile mutant strain (Figs.~\ref{fig:expts-1}B and \ref{fig:expts-1}D).
The onset pressure thus plays a significant role in regulating the mechanism by which bacterial monolayers transition out of plane. \smallskip


\section*{Motility-induced buckling of swarming colonies}
Having established a connection between cell density and the state of the colony (swarming vs. glassy in Figs.~\ref{fig:DF theory}B and \ref{fig:DF theory}C), we now consider the mechanisms leading to 3D growth of bacterial monolayers. 
Bacteria secrete extracellular polymeric substances that generate cell-substrate adhesion which keeps the monolayer stable against small perturbations. 
Sufficiently large destabilizing forces can overcome these adhesive forces and induce a 3D transition.
Above the onset cell density, bacteria enter into a glassy state with no significant flows within the monolayer.  
Here, the cell concentration increases slowly from division and growth, which generate larger in-plane interparticle stresses and induce buckling at a critical stress, similar to the classic Euler-Bernoulli beam bending instability, as observed previously \cite{Yan19,Beroz18,You19,Dell18,Grant14,Boyer11,Farrell13}.
As mentioned before, the characteristic timescale of buckling is significantly larger in the glassy state than in the swarming state; in fact, buckling can be described as quasi-static in the glassy state and for non-motile colonies.  
However, transitions into 3D in the swarming state below the onset density as observed in Fig.~\ref{fig:expts-1} have not been addressed, and their mechanisms are yet to be discovered, which we address below. \smallskip

An inspection of the swarming state of the monolayer shows a variety of motility induced in-plane flows resembling a 2D fluid, including linear flows such as simple shear, rotational, compressional and extensional flows, and nonlinear flows such as squeeze flows (see Fig.~\ref{fig:expts-1}B and SI movies S1-S3). 
In this case, the motility-induced velocity gradients can lead to in-plane shear stresses that destabilize the monolayer.
The ability of these in plane flows and the resulting fluid stresses leading to out-of-plane transitions can be understood by analyzing the following simple scenario. 
Let us consider a uniform velocity gradient that results in a simple shearing flow, i.e., Couette flow \cite{Bird07,leal2007advanced}. 
In Couette flow, the spatial gradient in velocity is determined by a shear strain rate $\dot{\gamma}$, which gives rise to compression and extension along two principal axes. 
For sufficiently large $\dot{\gamma}$, the forces along the compressional axis may enable the bacteria to overcome substrate adhesion forces, causing an out-of-plane deformation of the monolayer leading to ``buckling'', and thus nucleating the transition into the third dimension. \smallskip

The coupling between motility induced in-plane flows and out-of-plane deformations of the monolayer, and the onset of buckling, can be quantitatively analyzed by considering the monolayer as a 2D sheet, which can deform out-of-plane (see Fig.~\ref{fig:active-buckling-1}A and SI Chap.~III). 
Such couplings are common in other biological processes, in particular processes involving lipid bilayers \cite{seifert:1993, evans:1980, arroyo:2009, Sahu17} and biological membranes \cite{evans:1980}, which generally behavior as a fluid in plane and deform elastically in bending.
Inspired by the 2D theoretical framework involved in modeling lipid bilayers \cite{Sahu17,arroyo:2009,rangamani:2013}, we can write the out-of-plane equation of motion governing the height $h(\mathbf{x},t)$ of a fluid sheet connecting the mid-plane of bacteria subjected to various competing forces (Fig.~\ref{fig:active-buckling-1}A) as 
\begin{equation}\label{eq:out-of-plane}
\begin{split}
    \underbrace{\Delta p}_{\begin{subarray}{l}\text{bulk}\\
    \text{pressure}\end{subarray}} + &
    \underbrace{\lambda \nabla^2 h}_{\begin{subarray}{l}\text{tension-curvature}\\
    \text{coupling}\end{subarray}} + 
    \underbrace{\pi^{\alpha \beta} b_{\alpha \beta}}_{\begin{subarray}{l}\text{viscous-curvature}\\
    \text{coupling}\end{subarray}} = \\
    &  \underbrace{  \frac{1}{2} k_\text{b} \nabla^{2} \nabla^{2} h  }_{\begin{subarray}{l}\text{bending}\\ 
    \text{}\end{subarray}} + 
    \underbrace{E_{\text{ad}} h}_{\begin{subarray}{l}\text{cell-substrate}\\
    \text{adhesion}\end{subarray}} +
    \underbrace{\zeta_{\text{B}} \frac{\partial h}{\partial t}}_{\begin{subarray}{l}\text{bulk drag}\\
    \text{}\end{subarray}};
\end{split}
\end{equation}
see SI Chap.~III for a detailed derivation of the theory of 2D deformable fluid sheets. 
In \eqref{eq:out-of-plane}, $\Delta p$ is the bulk pressure acting on the monolayer, $\lambda=-\Pi(Pe)$ is the in-plane tension, $\pi^{\alpha \beta}$ are the in-plane viscous stresses, $b_{\alpha \beta} = h_{,\alpha \beta}$ is the curvature tensor, $k_\text{b}$ is the bending modulus, $E_{\text{ad}}$ is the strength of adhesion of the monolayer with respect to the substrate and $\zeta_\text{B}$ denotes a drag on the monolayer. \smallskip

Using the spherocylinder polydisperse system as appropriate for bacterial monolayers, the rheological behavior of the monolayer is found to be that of a non-Newtonian fluid (see SI Chap.~III.4.1). 
In this case, the in-plane viscous stresses $\pi^{\alpha \beta}$ in \eqref{eq:out-of-plane} are given by the non-Newtonian ``power law'' fluid constitutive equation \cite{Bird07}, $\pi^{\alpha \beta} = \eta_0 (v^{\alpha,\beta} + v^{\beta,\alpha})^n$, where $v^{\alpha,\beta}$ denote in-plane velocity gradients, $\eta_0$ is an effective viscosity, and $n$ is the power. 
This constitutive equation is particularly relevant for incompressible flows where $v^{\alpha, \alpha}= 0$, but we have also considered the case of compressible flows which will be discussed later.
Equation~(\ref{eq:out-of-plane}) is an extension of the Young-Laplace equation \cite{leal2007advanced} used to describe fluid films, but allows for additional mechanisms related to coupling of intra-membrane viscous flow and curvature $( \pi^{\alpha \beta} b_{\alpha \beta} )$, and cell-substrate adhesion that are relevant for understanding the growth of bacterial monolayers into the third dimension.   \smallskip

We now proceed to analyze the mechanisms and quantitative aspects of nucleation of colonies into the third dimension. 
In what follows, we begin by considering special cases of linear (uniform shear) and non-linear (squeeze) in-plane incompressible flows, linear compressible flows, and show how motility-induced flows lead to rate-dependent buckling instabilities. 
These special cases allow for exact analytical solutions, elucidate the physical mechanisms of out-of-plane transitions, and are further amenable to testing in both molecular simulations and experiments. \smallskip

\subsubsection*{\normalsize{Incompressible linear flows}}
We begin with uniform linear incompressible shear flows, i.e., Couette flows given by in-plane velocity gradients $\boldsymbol{\nabla}\mathbf{v} = [0,\dot{\gamma};0,0]$ (see Fig.~\ref{fig:active-buckling-1}B). 
The case of shear flow is particularly suitable for testing in molecular simulations, and plays a crucial role to interrogate our hypothesis of motility-induced buckling. In this case, we assume that the in-plane tension is negligible, \emph{i.e.,} $\lambda = 0$. 
Such a 2D flow corresponds to positive and negative eigenvalues, $\nu_1= \dot{\gamma}$ and $\nu_2= -\dot{\gamma}$, along two principal directions corresponding to the extensional and compressional axes, respectively. This indicates that the bacteria may buckle along the compressional axis for large $\dot{\gamma}$. 
To test this hypothesis, we perform linear stability analysis of \eqref{eq:out-of-plane} in terms of Fourier representation for the height, i.e., $h(\mathbf{x},t) =\sum_{\mathbf{k}} h_{\mathbf{k}} \exp(ik_\alpha x^\alpha +\omega t)$, where $h_{\mathbf{k}}$ is the Fourier component and $\mathbf{k}$ is the wave-vector, and $\omega(\mathbf{k})$ is the frequency (see SI Chap.~III.3).
Note that $\omega(\mathbf{k})>0$ indicates an unstable out-of-plane deformation, which ultimately leads to buckling of the surface. 
Figure~\ref{fig:active-buckling-1}B shows the dispersion curves $\omega(\mathbf{k})$ in the presence and absence of bending for various values of a non-dimensional parameter $\overline{E}_{\text{ad}} = {E_{\text{ad}} b^2}/({\eta_0 \dot{\gamma}^n})$, which describes the ratio of adhesive forces to viscous forces in Eq.~(\ref{eq:out-of-plane}) with $b$ being the bacterial length. 
As can be seen in Fig.~\ref{fig:active-buckling-1}B, when $\overline{E}_{\text{ad}}$ is high, $\omega(\mathbf{k}) <0$ for all wave-vectors, indicating that adhesive forces dominate the viscous-curvature coupling forces, and the monolayer is stable.
However, for sufficiently low $\overline{E}_{\text{ad}}$, i.e., when motility-induced viscous stresses are large, there exist wave-vectors where $\omega(\mathbf{k}) > 0$, which lead to buckling, and subsequent growth of the monolayer into the third dimension. \smallskip

In the absence of bending stiffness, the dispersion relation is given by $\omega \zeta_\text{B} = -2 \eta_0 k_1 k_2 \dot\gamma^n - E_{\text{ad}}$, which shows that unstable modes are possible whenever $k_1 k_2 < 0$ (Fig.~\ref{fig:active-buckling-1}B). 
Moreover, growth rate increases monotonically with wave-vector, so the fastest growth of deformation occurs at a wave-vector corresponding to the smallest length scale in the system---an individual bacterial size---indicating that buckling occurs at the level of individual bacteria.
This remarkable prediction is consistent with both our experiments and molecular simulations showing individual bacteria popping out into the third dimension in a spatially heterogeneous manner, as opposed to rafts of bacteria delaminating from the monolayer.  
Note that strong cell-cell adhesion can introduce a finite bending stiffness, which introduces a fourth-order stabilizing term, $\kappa_\text{b} (k_1^2 + k_2^2)^2$, leading to a maximum growth rate at intermediate wave-vectors. 
This can lead to rafts of bacteria that buckle like an elastic sheet as opposed to our experimental studies. 
We leave this possibility to future work. \smallskip

In summary, our linear stability analysis of shear flows (see SI Chap.~III.3.2) dictates that buckling occurs whenever 
\begin{equation} \label{eq:adhesion}
    \frac{1}{8 \pi^2} \frac{E_{\text{ad}} b^2}{\eta_0 \dot\gamma^n} \leq 1,
\end{equation}
indicating that growth into 3D occurs when adhesive and viscous-curvature coupling forces are comparable. 
We further analyzed other linear flows such as uniform rotational, and incompressible extensional flows (see SI Chap.~III.3), and find that rotational flows do not lead to any buckling instabilities. 
The latter case of incompressible extensional flows is similar to that of the shear flows with identical dispersion curves, and therefore similar mechanisms of buckling. \smallskip

\subsubsection*{\normalsize{Incompressible non-linear flows}}
In what follows, we consider a special case of incompressible nonlinear flow, hereinafter, referred to as the squeeze flow. 
In this case, we assume that the general constitutive behavior of the compressible fluid is that of a Newtonian fluid instead of a non-Newtonian fluid as considered previously for linear incompressible flows. 
Such a simplification allows for the possibility of analytical solutions demonstrating the role of non-uniform flows without altering the nature of the physical mechanisms leading to buckling. 
For an incompressible Newtonian fluid, the viscous stresses are given by $\pi^{\alpha \beta} = \eta_0 (v^{\alpha,\beta} + v^{\beta,\alpha})$, where $\eta_0$ is the Newtonian shear viscosity. 
In a natural swarm, it is possible that there exist two sufficiently large motile flocks of bacteria with mean velocity $U$ of a length scale $L$ travel towards each other, but can only disperse in a constricted manner in a gap $2\delta$ in the orthogonal direction where $\delta \ll L$; see Fig.~\ref{fig:active-buckling-1}C for a schematic description.
Such a scenario corresponds to a rate of squeezing $\dot{\gamma} \sim U/\delta$.
This generates compressive forces in the center of the squeeze flow region (see SI Chap.~III.3.4), which again leads to buckling and nucleation of growth in the third dimension.
This build up of the compressive stresses is simply due to inability of bacteria to escape through the constricted region $2\delta$ quickly enough.  \smallskip

Using lubrication theory in fluid systems for analyzing squeeze flows \cite{leal2007advanced} and performing the linear stability analysis yields dispersion curves $\omega(\mathbf{k})$ shown in Fig.~\ref{fig:active-buckling-1}C (see SI Chap.~III.3.4 for details). There exists one significant difference between the linear shear and non-linear squeeze flows. 
In the case of shear flows, buckling and growth in third dimension is a result of the competition between shear stresses $\pi^{\alpha \beta}$ and adhesive forces. 
However, for squeeze flows in a thin gap, the effects from negative tension $\lambda \sim -\eta_0 (U/\delta) (L/\delta)^2$ dominate the viscous stresses $\pi^{\alpha \beta} \sim \eta_0 (U/\delta) $ by a factor of $(L/\delta)^2 \gg 1$. 
This is reflected in the dispersion curves (Fig.~\ref{fig:active-buckling-1}C), where even for high values of $\overline{E}_{\text{ad}}$ when the monolayers are stable under linear shear flows (Fig.~\ref{fig:active-buckling-1}B), there exist $\omega(\mathbf{k}) > 0$ for squeeze flows leading to buckling of the monolayer.  
In this case, the linear stability analysis (see SI Chap.~III.3.4) dictates that buckling occurs whenever
\begin{equation}
    \frac{1}{2 \pi^2} \frac{ E_{\text{ad}} b^2}{\eta_0 \dot\gamma}  \Big(\frac{\delta}{L}\Big)^2 \leq 1,
\end{equation}
i.e., when the adhesive forces $({E}_{\text{ad}}h)$ and the flow induced compressive (or negative tensile) forces $(\lambda \nabla^2 h)$ resulting from squeezing effects are comparable in \eqref{eq:out-of-plane}. 
This case explicitly demonstrates the role of rate-dependent tension-curvature mediated instability, where in-plane flows result in gradients of tension. 
Such gradients in tension are dependent on flow characteristics and therefore are again motility induced.
Note that the growth rate $\omega(\mathbf{k})$ grows monotonically with wave-vector, as in the case of shear flows, leading again to buckling at the level of individual bacteria (Fig.~\ref{fig:active-buckling-1}C).\smallskip
 
\subsubsection*{\normalsize{Compressible linear flows}}
We now study buckling instabilities arising from compressible flows by considering uniform velocity gradients of the type $\boldsymbol{\nabla}\mathbf{v}= [\nu_1,0;0,\nu_2]$ (see SI Chap.~III.3.5 for a detailed analysis). 
The case of $\nu_1+\nu_2 \neq 0$ corresponds to compressible flows, while $\nu_1+\nu_2 = 0$ represents strictly incompressible flows as analyzed  previously. 
Since $\nu_1$ and $\nu_2$ are also the eigenvalues along the two principal directions and need not be equal, this choice of the velocity gradient includes all of the aforementioned uniform linear flows including incompressible shear, rotational and extensional flows. 
When considering compressible flows, the in-plane viscous stresses $\pi^{\alpha \beta}$ will also contain additional contributions from bulk viscosity. 
Here again, we assume that the general constitutive behavior of the compressible fluid is that of a Newtonian fluid, which allows us to obtain simple analytical solutions that capture the role of compressible flows and bulk viscosity in buckling.
The viscous stresses for a compressible Newtonian fluid are given by $\pi^{\alpha \beta} = \eta_0 (v^{\alpha,\beta} + v^{\beta,\alpha}) +2\xi v^{\delta,\delta} \delta^{\alpha \beta}$, where $\eta_0$ and $\xi$ are the shear and bulk viscosities. 
This modifies the viscous-curvature coupling term $\pi^{\alpha \beta}b_{\alpha \beta}$ in \eqref{eq:out-of-plane} with an additional contribution from the bulk viscosity (see Fig.~\ref{fig:compression-buckling}).
To understand the buckling mechanisms arising from strictly compressible flows, we analyze two special cases corresponding to radially symmetric compression ($\nu_1 = \nu_2 = -\dot{\gamma}$) and extensile ($\nu_1 = \nu_2 = \dot{\gamma}$) flows.
We also analyze various cases of base tension $\lambda = - \Pi$, where the base pressure $\Pi$ is a function of activity $Pe$ and can be non-zero for large activities. 
Specifically, we find that the pressure increases with $Pe$ (see SI Fig.~2.18) and plays a significant role in influencing the nature of the buckling mechanisms, particularly for extensile flows.\smallskip
 
Figure~\ref{fig:compression-buckling}A shows the dispersion curves $\omega(\mathbf{k})$ arising from a linear stability analysis of the radial compression flow ($\nu_1=\nu_2=-\dot{\gamma}$) for different base pressures $\Pi$ and adhesion strengths (see SI Chap.~III.3.5). 
They are qualitatively similar to the case of uniform shear and non-uniform squeeze flows, where $\omega(\mathbf{k})$ again increases monotonically with wave-vector $\mathbf{k}$ leading to buckling at the length scale of individual bacteria. 
However, the fluid sheet buckles at even large adhesion strengths in comparison to shear flows (see Fig~\ref{fig:active-buckling-1}B). 
Furthermore, Fig~\ref{fig:compression-buckling}B shows that increasing the base pressure $\Pi$, corresponding to an increase in activity $Pe$, amplifies the buckling instability and makes the fluid sheet unstable to  larger adhesion strengths compared to the case $\Pi\approx 0$. 
In summary, the linear stability analysis dictates that buckling occurs whenever 
\begin{equation}
    \frac{1}{4\pi^2}\frac{E_{\mathrm{ad}}b^2}{\eta_0 \dot{\gamma}} \leq \Big( \frac{\Pi}{\eta_0 \dot{\gamma}} +  \big(
    4 \bar{\xi} + 3\big) \Big), 
\end{equation}
where $\bar{\xi} = \xi/\eta_0$ denotes the ratio of shear and bulk viscosities. \smallskip

We now consider radial extensile flows of the type $\nu_1 = \nu_2 = \dot{\gamma}$. Linear stability analysis yields the dispersion curves shown in Figure~\ref{fig:compression-buckling}B for different pressures and adhesion strengths (see SI Chap.~III.3.5). 
In this case, it can be seen that when there exists no base pressure, i.e. $\Pi=0$, the extensile flows are always stable as expected. 
However, when $\Pi$ is large, the growth rate $\omega(\mathbf{k})$ becomes positive leading to buckling even under extensile flows. The condition for which buckling occurs in the extensile case is given by 
\begin{equation}\label{eq:buckle-extensile}
    \frac{1}{4\pi^2}\frac{E_{\mathrm{ad}}b^2}{\eta_0 \dot{\gamma}} \leq 
    \Big( \frac{\Pi}{\eta_0 \dot{\gamma}} -  \big(4\bar{\xi} + 3\big) \Big). 
\end{equation}
Using \eqref{eq:buckle-extensile}, one can also see that the maximum extensile flow below which buckling one can observe buckling is given by 
\begin{equation}
    \dot{\gamma}_{\mathrm{max}} = \frac{\Pi - E_{\mathrm{ad}} b^2/(4 \pi^2) }{(4\xi + 3\eta_0)}.
\end{equation}
As we will show later, this counter intuitive condition of buckling under extensile flows is relevant in situations at higher activities and also in our experiments of \emph{P. aeruginosa} colonies.  

\subsubsection*{\normalsize{Molecular simulations}}
We now test our analytical theory of flow-induced buckling instabilities and out-of-plane growth in molecular simulations consisting of spherocylinders with polydispersity as of that of the realistic bacterial colonies. 
As mentioned before, the easiest scenario to test the hypothesis corresponds to the case of a non-motile monolayer subjected to in-plane simple shear rates. 
In this case, one can rigorously test the buckling instability relation derived in \eqref{eq:adhesion} comparing the adhesion forces and shear stresses. 
To this end, we consider a monolayer of polydisperse spherocylinders at varying external shear rates and substrate adhesion; see Fig.~\ref{fig:molecular-testing}A, SI movies S10-S11 and SI Chap.~III.4 for a detailed analysis.  
We observe that large shear rates cause out-of-plane transitions, defined as the moment a particle rises 1 particle diameter above the monolayer (shown in white in Fig.~\ref{fig:molecular-testing}A).  
Quantifying the instability across various adhesion and particle densities, the data show a clear relation separating the stable and unstable regions; see Fig.~\ref{fig:molecular-testing}B.
Remarkably, the data show excellent agreement for critical buckling given by the relation $E_{\text{ad}} \sim \eta_0 \dot\gamma^{0.5}$; see Fig.~\ref{fig:molecular-testing}B.  
Upon conducting further simulations to obtain the stress-strain relationship for the same polydisperse spherocylinder system (see SI Chap.~III.4), we find the monolayer to be a power-law, shear-thinning fluid with an effective viscosity $\eta^{\text{eff}} \sim \dot\gamma^{-0.5}$, consistent with dense colloidal suspensions \cite{Isa07, Petekidis04, Lin18}. 
This then yields the adhesion-viscosity relation, $E_{\text{ad}} \sim \eta^{\text{eff}}(\dot\gamma) \dot\gamma = \eta_0 \dot\gamma^{0.5}$ in \eqref{eq:adhesion}.
Furthermore, Fig.~\ref{fig:molecular-testing}B also shows that all the data collapse on to single point when scaled by \eqref{eq:adhesion}, thus confirming the mechanism of shear flow induced buckling instability. 
Lastly, we observe that buckling always occurs at the level of individual particles (i.e., largest possible wave vector), consistent with both the analytical theory and experimental observations.\smallskip

Molecular simulations corresponding to linear shear flows demonstrate the aspect of flow-induced growth of bacterial colonies into the third dimension. 
We now test the applicability of our proposed buckling mechanisms to an increasingly complex case of non-uniform flows in molecular simulations, corresponding to localized shear flows. 
To this end, we consider two adjacent flocks that are driven uniformly in opposite directions as shown in Fig.~\ref{fig:molecular-testing}C (see SI movie S12). 
This is reminiscent of two flocks moving in opposite directions in realistic bacterial colonies. 
In this case, velocity gradients are localized to a boundary layer around the interfaces where the two flocks meet, and there is one negative eigenvalue $\nu_{\text{min}}$ that indicates compression at the interfaces (see SI Chap.~IV.1 for a description of eigenvalue analysis). 
These negative eigenvalues are shown as a contour plot in Fig.~\ref{fig:molecular-testing}D, with corresponding flow fields in red arrows (see SI movie S13).  
If the aforementioned buckling instabilities arising from the shearing flows is true, one should observe buckling only at the interfaces between two flocks due to compression. 
This is evident in our simulations in Fig.~\ref{fig:molecular-testing}E, where buckling is localized to the interfaces where the two flocks meet, thus providing credence on the generic applicability of our motility-induced instability mechanisms. 

\section*{Experiments vs. simulations}
We now study the applicability of our motility induced buckling mechanism in real bacterial swarms, which contain a variety of spatially heterogeneous in-plane flow fields, in addition to homogeneous linear flows and inhomogeneous nonlinear flows. 
The inherent activity of the bacteria produces natural swarms, and the system will experience a wide range of spatially and temporally-varying shear, tensile, and compressive stresses, which require a new metric to correlate the flow fields to buckling events. 
To this end, we calculate the rate-of-strain tensor at each fluid element, in both experiments on bacterial colonies and molecular simulations of active polydisperse spherocylinder systems, defined by $\mathbf{E} = (\nabla \mathbf{v} + (\nabla \mathbf{v})^T)/2$. The strain-rate tensor contains two eigenvalues $\nu_{\mathrm{min}}$ and $\nu_{\mathrm{max}}$. 
A negative and positive eigenvalue indicates compression and extension, respectively, in a direction informed by the corresponding eigenvectors. 
Assuming that buckling is correlated by an appearance of a bacterium onto the second layer, we note the corresponding position and the associated eigenvalues in the bottom monolayer just prior to buckling. 
These eigenvalues can be represented as in Fig.~\ref{fig:eigenvalue-diagram}A, with the smaller value $\nu_{\text{min}}$ on the horizontal axis, and the larger value $\nu_{\text{max}}$ on the vertical axis. \smallskip
 
The eigenvalue diagram in Fig.~\ref{fig:eigenvalue-diagram}A can be partitioned into three distinct zones, each marking a different motility-induced buckling mechanism:
(I) Incompressible shear flow ($\nu_{\text{min}} + \nu_{\text{max}} \approx 0$), in pink;
(II) Compression and shear flow ($\nu_{\text{min}} + \nu_{\text{max}} < 0$), in green;
(III) Extensile flow ($\nu_{\text{min}} + \nu_{\text{max}} > 0$), in blue.  \smallskip

We first test the utility of the metric devised in Fig.~\ref{fig:eigenvalue-diagram}A on describing the buckling mechanisms found in MD simulations of localized shear flows shown in Figs.~\ref{fig:molecular-testing}C-E. 
Since there exists uniform shear stress along the interface between the two flocks, we expect the buckling events to correspond to zone (I), and to lie on the line $\nu_{\text{min}} + \nu_{\text{max}} \approx 0$.  
As expected, Fig.~\ref{fig:eigenvalue-diagram}B shows perfect agreement with shear flow mediated buckling for a range of external forcing. Figure.~\ref{fig:eigenvalue-diagram}B shows the population of buckled particles corresponding to distinct zones again demonstrating that buckling occurs from incompressible shear flow as expected. 
The case of homogeneous shear flow shows similar results (see SI Fig.~4.2).\smallskip

We now apply our metric to experiments and simulations with active swarms to predict motility-induced out-of-plane colony growth. 
Figures~\ref{fig:exps-sims}A and \ref{fig:exps-sims}B summarize the results of both MD simulations of active (or motile), polydisperse spherocylinders for various activities and experiments of bacterial colonies, respectively, for densities lesser than the onset density for glassy dynamics. See SI Chap.~IV for details on MD simulations of active spherocylinders and experimental quantification. 
Figure~\ref{fig:exps-sims} (top row) shows snapshots of molecular simulations and experiments, where the red arrows are the velocity field, $\vb(\xb,t)$, the color contour indicates $\nu_{\text{min}}$, and the white circular markers indicate cells that buckled at this time step (see SI movies S14-S16). 
Note that our MD simulations exhibit swarming flows, consistent with experimental colonies in Fig.~\ref{fig:exps-sims}B (top). 
Smaller activities ($Pe=5$ in Fig.~\ref{fig:exps-sims}A) produce buckling events corresponding to shear flows with rates that are in the vicinity of the unstable region in Fig.~\ref{fig:molecular-testing}B. 
Careful examination of these events reveals that buckling occurs frequently at interfaces between two opposing flows with localized shear stress, similar to interface flows studied in Fig.~\ref{fig:molecular-testing}C-E.
In this case, we expect incompressible shear flow mediated buckling, corresponding to zone (I), as confirmed in Fig.~\ref{fig:exps-sims}A (middle and bottom), indicating that buckling events are triggered by a competition of the viscous-curvature coupling term ($\pi^{\alpha \beta} b_{\alpha \beta}$) and cell-substrate adhesion ($E_{\text{ad}} h$) consistent with the instability relation in \eqref{eq:adhesion}. \smallskip 

As activity increases, swarming patterns evolve more quickly in time and we obtain nonlinear and/or compressible flows, including squeeze and radial compression/extensile flows. 
This changes the out-of-plane growth from a simple shear flow induced buckling mechanism (zone I) to that of compression, shear and squeeze flow  (zone II). 
Interestingly, we also observe buckling in extensile flow regions (zone III), which may be counter-intuitive based on a sole competition between adhesion $E_{\text{ad}}$ versus viscous-curvature coupling $\pi^{\alpha \beta}$.
These results indicate the presence of a nonzero in-plane active pressure $-\lambda = \Pi(Pe,\phi)$ and a significant role of tension-curvature coupling in \eqref{eq:out-of-plane}, as previously analyzed for radial extensile flows leading to \eqref{eq:buckle-extensile}.
In this case, we see that swarming colonies can buckle even in purely extensile regions, if the in-plane pressure is large enough to overcome the stabilizing effects of the extensile flows, as explained in Fig.~\ref{fig:compression-buckling}B.
Ultimately, it is the competition between $E_{\text{ad}}$ (always stabilizing), $\lambda$ (always destabilizing), and $\pi^{\alpha \beta}$ (either stabilizing or destabilizing depending on flow profile).
Indeed, eigenvalue plots and histograms for $Pe=7$ show transitioning of buckling events from zone (I) to zones (II) and (III). 
For $Pe=10$, the buckling events further spread uniformly across all zones to shear, squeeze, and radial compressible flows, as shown in Fig.~\ref{fig:exps-sims} (middle and bottom rows). 
Therefore, if motility is strong compared to cell-substrate adhesion, we predict buckling events via a combination of mechanisms involving both viscous-curvature and tension-curvature coupling arising from shear, squeeze, and radial flows. \smallskip
 
Figure~\ref{fig:exps-sims}B shows the analysis of buckling events in motile \textit{P. aeruginosa} swarms at densities smaller than the onset density for glassy dynamics in Fig.~\ref{fig:DF theory}C (see SI movies S17-S18). 
Here, we observe the buckling events spread across all zones (Fig.~\ref{fig:exps-sims}B (middle and bottom)), indicating that swarming \textit{P. aeruginosa} colonies exhibit all types of flow characteristics, 
similar to our simulations at intermediate activity $7 < Pe < 10$.  
In fact, for the twitching motility speed of $U_0 \sim 10 \mu \text{m/min}$, reorientation time of $\tau_\text{R} \sim 1 \text{min}$, and characteristic bacteria size of $\sigma \sim 1 \mu \text{m}$, the activity of the bacteria corresponds to $Pe = U_0 \tau_\text{R} / \sigma \sim 10$. 
\textit{P. aeruginosa} are known to secrete extracellular polymeric substances that facilitate cell-cell adhesion \cite{Lyczak00,Gellatly13,Faure18,Lopez15,Bjarnsholt13}; this may partially stabilize monolayers under extensile flows and help explain the lower population in zone (III) compared to the MD simulations, which do not incorporate cell-cell adhesion. 
Although cell-substrate and cell-cell adhesion cannot be calculated easily in experiments, an activity of $Pe \approx 7-10$ gives results consistent with those of our simulations, where buckling is triggered by all mechanisms in \eqref{eq:out-of-plane}. \smallskip

Taken together, Figs.~\ref{fig:exps-sims}A and \ref{fig:exps-sims}B show that motility-induced buckling mechanisms mediate out-of-plane colony growth.
In summary, at densities lower than the onset density, motile bacterial colonies exist in a swarming state and generate motility-induced flows resulting in viscous and compressive stresses. 
These stresses then induce out-of-plane deformations, subsequently leading to buckling and growth into the third dimension. 
At cell densities above the onset density, the colony enters into a kinetically-arrested glassy state that suppresses any in-plane flows, and the nature of buckling crosses over from flow-mediated buckling to a quasi-static growth and division-induced Euler-Bernoulli buckling \cite{Yan19,Beroz18,You19,Dell18,Grant14,Boyer11,Farrell13}.  
Our dynamical state diagram in Fig.~\ref{fig:DF theory}C thus provides a fundamental framework to understand and regulate the 3D transition of bacterial colonies.

\section*{Discussions}
Recent works \cite{Saw17,Kawaguchi17} have focused on topological defects in motile mammalian tissues and their role in ``extrusion'' or buckling out of a monolayer, by considering the tissue to be an active nematic.  
For motile bacterial colonies of \textit{P. aeruginosa}, we observe no indication of out-of-plane transitions occurring preferentially at sites of defects (see SI Chap.~III.5 and SI movies S19-S20).  
Two large differences that contribute to this: (i) bacterial motility speeds ($\sim \mathcal{O}(10 \mu \text{m/min})$) are much larger than that of mammalian cells in \cite{Saw17,Kawaguchi17} ($\sim \mathcal{O}(10 \mu \text{m/hr}) $), and (ii) the aspect ratio of bacterial body is small ($\ell/\sigma \sim 3-4$) with significant polydispersity. 
At these aspect ratios, equilibrium phase diagrams show that monodisperse spherocylinders exist either in isotropic or smectic phases \cite{Bolhuis97,Bates00,Bautista14}.
We observe that adding polydispersity in our MD simulations disrupts the smectic phases and results in a micro-crystalline smectic structure. 
Furthermore, adding activity to MD simulations yields a flowing polycrystal (i.e., small domains of flocks), and not an active nematic; this behavior is also observed in swarming \textit{P. aeruginosa} colonies.
In this case, individual cells and crystals can reorient readily and transient polar flocks are disrupted more easily, thus making defects irrelevant to out-of-plane transitions in this study.
We also performed MD simulations of larger aspect ratio particles ($\ell/\sigma = 10$) and found that in these cases stable defects form, and are consistent with defects present in motile microtubules \cite{DeCamp15} and other slender biopolymers. 
At these high aspect ratios, we observe active nematic phases with long range orientational order and associated defects; see SI Chap.~III.5 for further discussion of defects, polycrystallinity and role of active nematics. 
It is of interest to consider motile bacteria with larger aspect ratios, such as Myxobacteria swarms, where the coupling between topological defects arising in active nematics and motility induced in-plane flows may play a significant role in the out-of-plane growth. Such a correlation between the defects and growth has been observed recently in myxobacterial swarms \cite{copenhagen2020topological}. 

Our theoretical and experimental analyses predict only the onset of buckling, or a nucleation event. 
In addition to 3D transitions, we have also observed a rich diversity of swarming behaviors in \textit{P. aeruginosa} colonies, including patterns at the edges of colonies that resemble fingering instabilities \cite{Gloag13,Verstraeten08}. 
It has been known that extracellular DNA is critical for biofilm formation \cite{whitchurch2002extracellular} and its degradation reduces swarming in \textit{P. aeruginosa} colonies \cite{Gloag13}.
We experimentally observe that the addition of DNA-degrading enzyme DNaseI suppresses swarming and subsequent buckling at leading edges (see SI Chap.~IV.3 and SI movie S21-S22).
These observations further support our motility-induced buckling mechanism; any perturbation that eliminates swarming prevents rate-dependent buckling. 
Other biochemical factors that impact the developmental program of bacterial colony growth require reevaluation of the dynamical state diagram (in Fig.~\ref{fig:DF theory}C), and consequently the motility-induced buckling mechanisms.   

Our work demonstrates that motile bacterial colonies can exist in different phases (liquid, glassy, active nematic, etc.), and that their properties in such states are relevant to the dynamical development of a colony. Such a scenario is relevant not only for bacterial colonies but may be broadly applicable to a wide variety of living systems such as yeast \cite{Atis19} and confluent mammalian cells \cite{Angelini11,Saw17,Kawaguchi17}.

\section*{Methods}

\subsubsection*{\textit{P. aeruginosa} cell culture and under-agar assay} 
Standard cell culture techniques were used to handle \textit{P. aeruginosa} strain PAO1.  
Briefly, PAO1 expressing GFP from a pSMC2 plasmid was grown in Luria broth (LB) nutrient media with $100 \mu \text{g/mL}$ carbenicillin.  
An overnight culture was diluted into fresh media and grown for 2 hours before centrifuging down to a desired concentration.
For the under-agar assay, a small volume ($\approx 3 \mu L$) of cell suspension was carefully placed on a cut agar nutrient pad containing LB media, $1.5 \%$ agarose (IBI Scientific Molecular Biology Grade), $100 \mu \text{g/mL}$ carbenicillin. 
For DNase experiments, $3 \mu$M SYTOX Orange Nucleic Acid Stain (Invitrogen, S11368) and 250 Kunitz/mL DNaseI from bovine pancreas (Sigma Aldrich, D5025) were additionally added within the agar pad.
The drop of cell suspension was left upright to air dry the liquid.  
A glass-bottom 35 mm dish, No.~1.5 Coverslip, (MatTek corporation) was carefully placed on top of the cut agar pad to create a thin layer of cells between the glass and the agar pad as shown in Fig.~\ref{fig:expts-1}A.
The chamber was sealed with parafilm and incubated in $37^\circ$C for 1-2 hours before imaging.
Experiments using \textit{P. aeruginosa} strain PA14 were also conducted and showed similar behavior as PAO1.  

Confocal imaging was performed using a Nikon Eclipse Ti microscope (Nikon Instruments) with a Yokogawa CSU-X spinning disk and an Apo TIRF 100x oil-immersion objective (numerical aperture $NA=1.49$). 
Two solid state lasers at wavelengths 488nm and 561nm  were used for excitation (ILE-400 multi-mode fiber with BCU, Andor Technologies), and filtered with emission filters (535/40m and 610/75, Chroma Technology). Experiments were also conducted with bright field illumination only, which indicated no change in out-of-plane transition or glassy behavior.  Photodamage of cells does not affect our results.

Images were acquired on a Zyla 4.2 sCMOS camera (Andor Technologies), using open-source Micromanager software.  
Different $(x^1,x^2)$ positions of the colony were scanned, and only those that began with a single monolayer were imaged.
Four planes of z-slices were acquired every $4-60$ sec, depending on the system.  
ImageJ (NIH) and MATLAB (Mathworks, Inc) were used to analyze data from images.
Open-source software SuperSegger \cite{Stylianidou16} and PIVLab \cite{Thielicke14} were used to track cell positions and velocities.
 
\subsubsection*{Molecular simulations} 
To explore a molecular model of motile bacterial colonies, we performed coarse-grained molecular dynamics (MD) simulations using the active Brownian particle model.  
See SI for a full description of the simulations.
Briefly, the overdamped Langevin equations with an active propulsive force was added to each particle as a body force.  
Hydrodynamic interactions were not included in our model.
We varied the area fraction $\phi$ and the activity parameter given by the P\'eclet number $Pe = U_0 \tau_{\text{R}}/\sigma$, a non-dimensional number defined as the ratio of the intrinsic run length of the active particle $U_0 \tau_{\text{R}}$ over its diameter $\sigma$.  
 
\begin{figure*}[t!]
\centering
\includegraphics[scale=0.61]{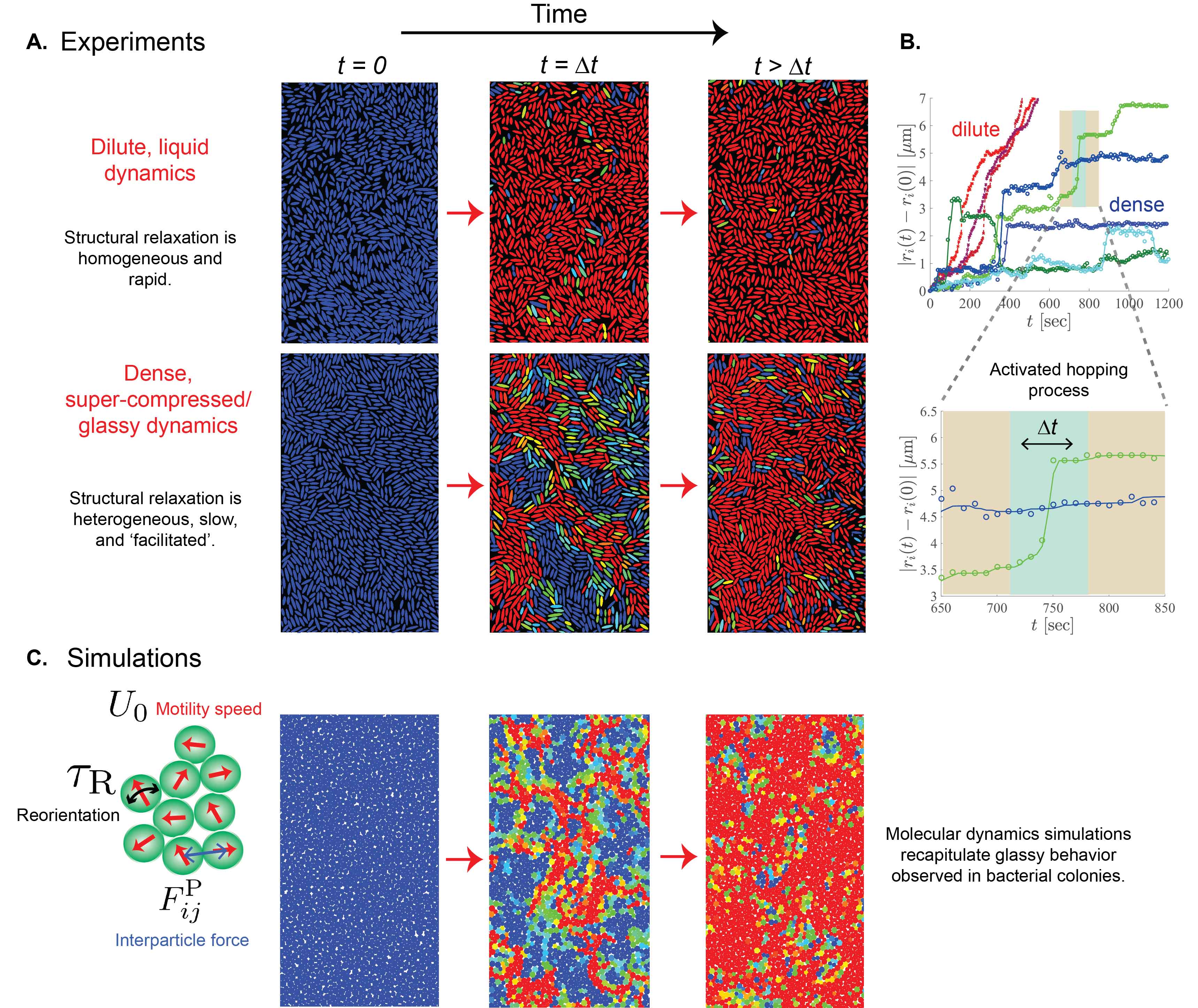}
\caption{Bacterial colonies exhibit signatures of glassy dynamics at large cell densities.  (A) Snapshots of experiments for dilute (top) and dense (bottom) bacterial colonies, taken at three successive time points.  Colors indicate displacement magnitude of each bacterium from its initial position, where blue indicates zero displacement and red indicates displacement more than a body length, $2\mu \text{m}$. The dilute colony shows immediate relaxation of the system, where all bacteria are displaced at least a body length in a period of time $\Delta t$ (defined later), as indicated by the rapid blue to red transition. In contrast, a dense bacterial colony exhibits dynamic heterogeneity, a characteristic signature of glassy dynamics, which results in distinct mobile and immobile regions in time $\Delta t$, as apparent by the strings of red that connect across large lengths scales. Here $\Delta t$ is the instanton time, i.e. the average time scale associated with a bacteria translating a body length as seen in (B). (B) Absolute values of particle displacement, $|\rb_i(t) - \rb_i(0)|$, corresponding to trajectories of several cells in both dilute and dense colonies. A magnified view of the trajectory is shown where a hopping event, similar to activated hopping events in molecular and colloidal glassy systems, is observed, thus exhibiting another characteristic signature of glassy dynamics.  (C) Left: A Schematic showing the molecular model and key parameters in our molecular dynamics (MD) simulations, where we use a 2D active Brownian particle model with motility speed $U_0$, reorientation time $\tau_{\text{R}}$, bidispersity ratio of $1.4$, and interparticle force $F_{ij}^{\text{P}}$ (see SI Chap~II.2.1 for details of MD simulations).  Right: Snapshots of simulations at three successive time steps for a dense particle monolayer exhibiting glassy dynamics. MD simulations of active Brownian particles at high densities recapitulate the dynamic heterogeneity, activated hopping events and other signatures of glassy dynamics observed in experiments.}
\label{fig:expts-glass}
\end{figure*}

\begin{figure*}[t!]
\centering
\includegraphics[scale=0.62]{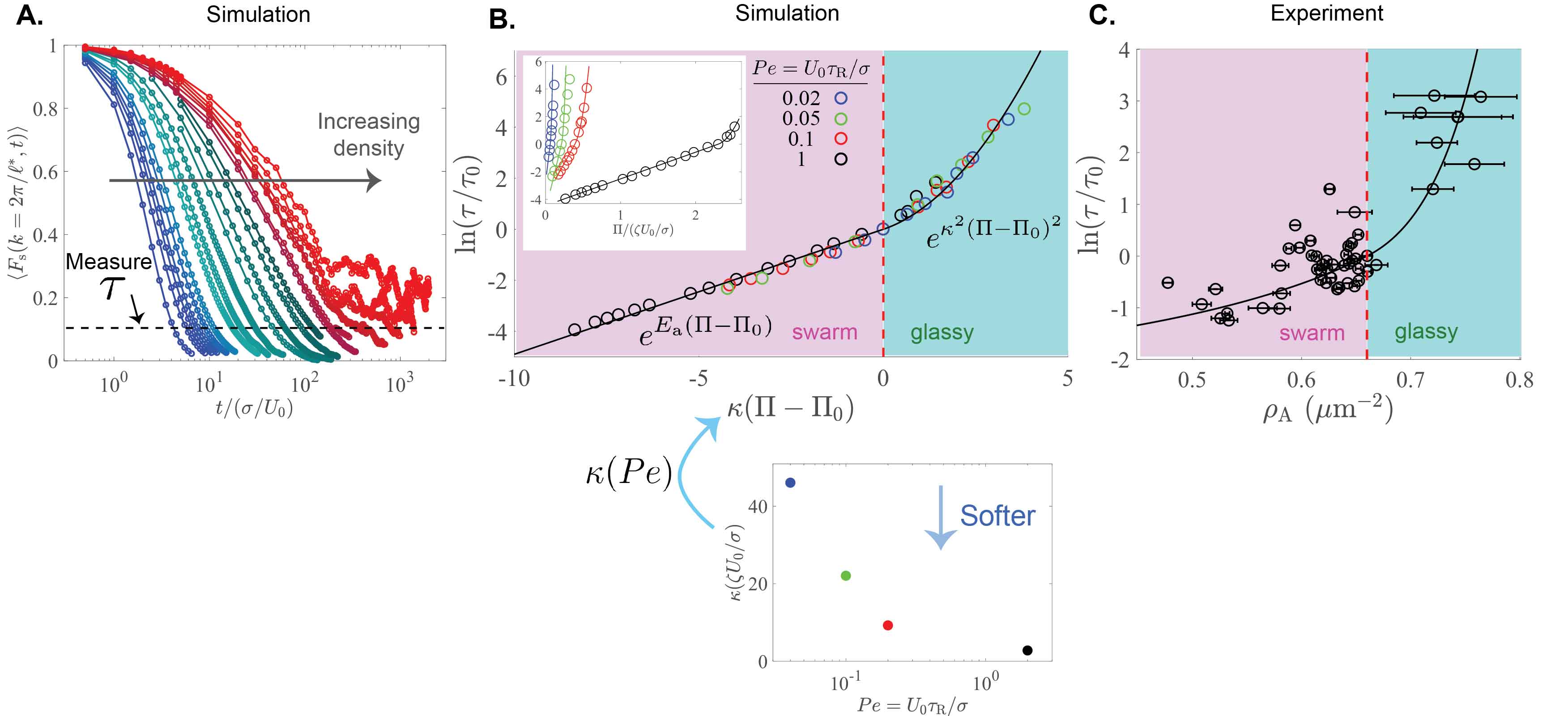}
\caption{Bacterial colonies behave as active glassy materials and can be described by an active dynamical facilitation (DF) theory.  
(A) Dynamic structure factor of motile, bidisperse particles in MD simulations across different densities, $\langle F_{\text{s}}(k = 2\pi/\ell^*,t) \rangle$, where $\ell^*$ is the length associated with the first peak in the static structure factor, $S(k)$ (see SI Fig.~2.9). 
Relaxation times from MD simulations and of the bacterial colonies are determined from the decay of $\langle F_{\text{s}}(k = 2\pi/\ell^*,t = \tau) \rangle$.
(B) Relaxation times from MD simulations as a function of mechanical pressure $\Pi$, for various values of activity $Pe=({U_0 \tau_\mathrm{R}})/{\sigma}$, show crossover from Arrhenius-like behaviors to super-Arrhenius behaviors at high densities, similar to the behaviors characteristic of molecular and colloidal glassy systems.  
All data (inset) collapse universally onto the parabolic law, $\kappa^2(\Pi - \Pi_0)^2$, as predicted by the DF theory, where $\kappa$ is related to the energy barrier corresponding to the emergence of excitations, and $\Pi_0$ is the onset pressure above which glassy dynamics is observed. The onset relaxation time $\tau_0$, energy barrier $\kappa$, and onset pressure $\Pi_0$ are a function of activity $Pe$.  
Below: Energy barrier $\kappa$ decreases with increasing $Pe$, where larger motility makes the material effectively `softer' with a shorter relaxation time, given the same density (or pressure).  
(C) Relaxation time from experiments as a function of cell density; the solid curve is our prediction from our extended DF theory, where an equation of state was used to convert between cell density and pressure (see SI Chap~II.4). The onset density or the onset pressure demarcates the swarming and glassy states of the colony.}
\label{fig:DF theory}
\end{figure*}

\begin{figure*}[t!]
\centering
\includegraphics[scale=0.52]{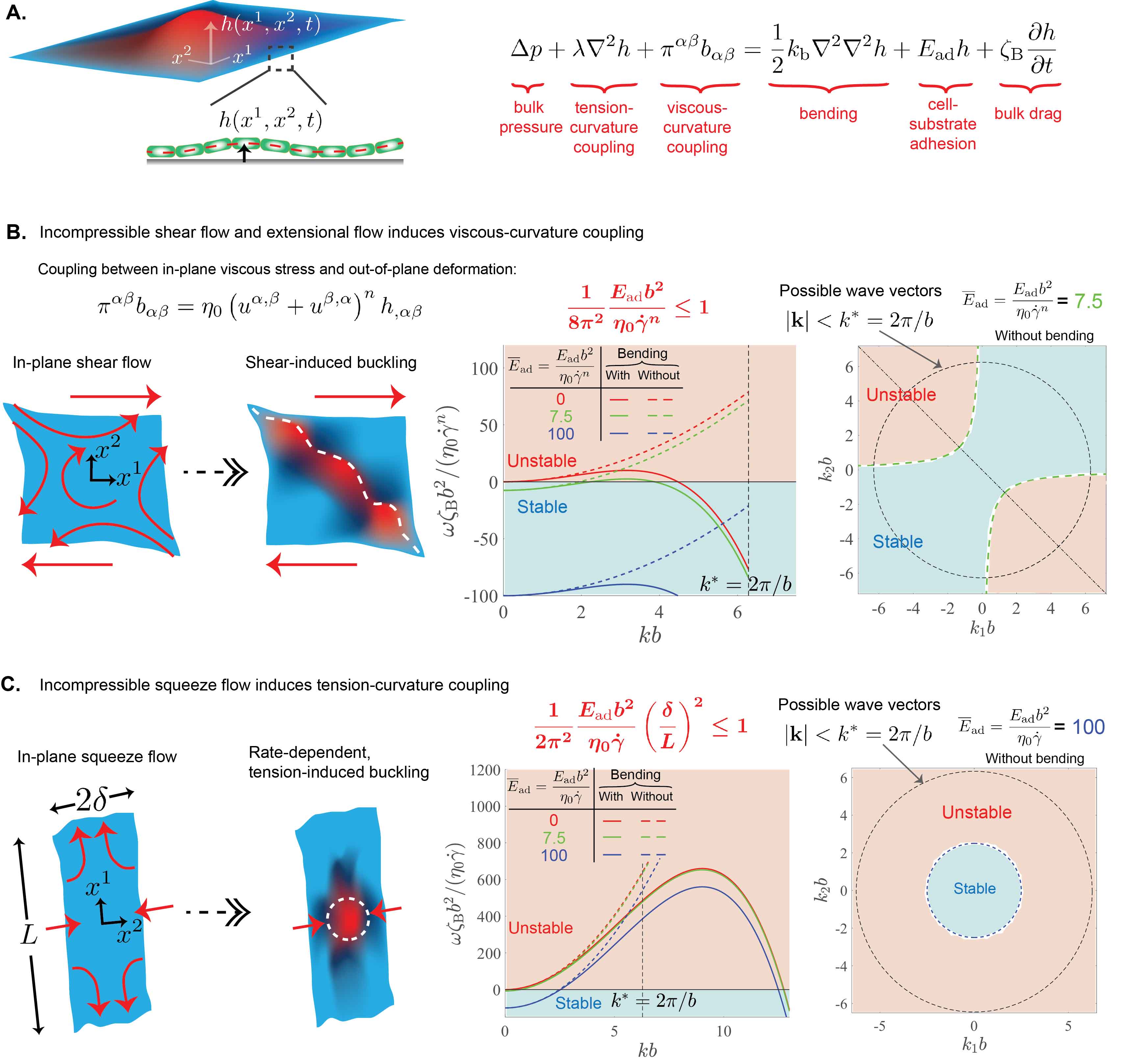}
\caption{
Proposed mechanisms of motility-induced buckling.  
(A) Left: A schematic of a collection of bacteria modeled as a 2D fluid film connecting the mid-plane of the bacteria, where $h(x^1,x^2,t)$ is the height of the mid-plane. 
Red regions indicate vertical displacement of the monolayer. 
Right: Out-of-plane momentum balance of a thin monolayer describing the height deformations of the bacterial film. 
The new term in this work corresponds to the viscous-curvature coupling $\pi^{\alpha \beta} b_{\alpha \beta}$, which relates in-plane viscous stresses to out-of-plane deformations. 
(B) Left: A schematic of in-plane incompressible shear flows inducing buckling along the compressional axis via viscous-curvature coupling. 
The white dashed lines indicate the undulations of the monolayer that subsequently lead to nucleation of individual cells out of plane.  
Center: Stability diagram showing the growth rate of unstable modes as a function of wave vector along the compressional axis, $k_1 = -k_2 = k$, for various cell-substrate adhesion $E_{\mathrm{ad}}$ and shearing rates $\dot{\gamma}$, in the presence (solid curves) and absence (dashed curves) of bending, where  $\omega(k) > 0$ indicate unstable regions. 
We anticipate that bending plays a small role compared to cell-substrate adhesion for bacterial colonies. 
In the absence of bending, the dispersion relation increases monotonically with $k$, and $\omega(k)$ is maximum at the largest wave vector, $k^* = 2\pi/b$, where $b$ is the bacterial length. 
This indicates that buckling from in-plane shearing flows occur at the level of individual bacteria.  
Right: Stability diagram in the full $k_1-k_2$ plane with the onset of instability demarcated by a dashed green line, for the case without bending and a fixed nondimensional adhesion $\overline{E}_{\mathrm{ad}} = 7.5$.
The dashed black circle indicates the maximum possible wave vector in our system, $k^* = 2\pi/b$.
(C) Left: A schematic of in-plane incompressible squeeze flow arising from flocks of bacteria of length $L$ moving towards each other but can only disperese over a length scale $H \ll L$.  
The red region indicates a stagnation point, leading to a build up of compression, which leads to buckling by tension-curvature coupling term $\lambda \nabla^2 h$ in the momentum balance.  
Center: Stability diagram for the case of squeeze flow, which leads to a maximum growth rate $\omega$ at $k^*=2\pi/b$ in the absence of bending, thus leading to buckling again at the level of individual bacteria.
Right: Stability diagram in the $k_1-k_2$ plane with the onset of instability demarcated by a dashed blue line, for the case without bending and a fixed nondimensional adhesion. $\overline{E}_{\mathrm{ad}} = 100$.
}  
\label{fig:active-buckling-1}
\end{figure*}

\begin{figure*}[t!]
\centering
\includegraphics[scale=0.60]{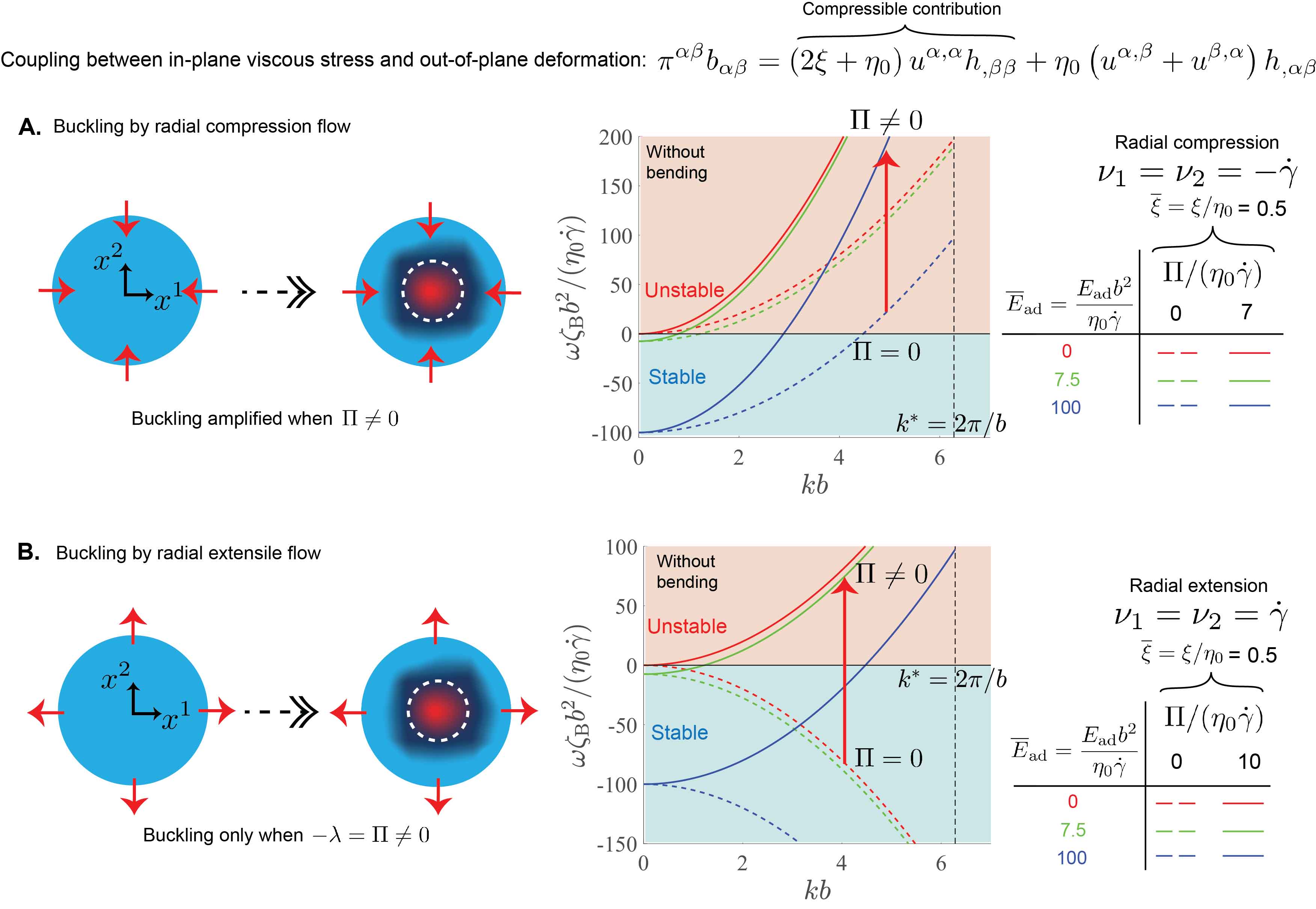}
\caption{
Proposed mechanisms of motility-induced buckling for the case of compressible flows. For compressible flows, there is an additional term involving the bulk modulus $\xi$ in the viscous stress tensor that may lead to buckling. 
(A) Left: Schematic of in-plane radial compression flow arising from flocks of bacteria converging to a single point.
Right: Stability diagram in the absence of bending, for a fixed bulk modulus $\xi/\eta_0 = 1$ and a range of cell-substrate adhesion. 
Nonzero bulk modulus amplifies the growth rate of all unstable modes.
In addition to compressibility, a nonzero in-plane pressure $\Pi$ amplifies the growth rate.  
As defined before, maximum growth rate occurs at the largest possible wave vector, $k^* = \frac{2\pi}{b}$, where $b$ is the bacterial length.
(B) Left: Schematic of in-plane radial extensile flow arising from flocks of bacteria migrating away from a single point.
Right: Stability diagram in the absence of bending, for a fixed bulk modulus $\xi/\eta_0 = 1$ and a range of cell-substrate adhesion. 
In the absence of in-plane pressure $\Pi$, extensile flows are stable across all modes. 
However, a sufficiently large in-plane pressure $\Pi$ can overcome the stabilizing effects of the extensile flow and lead to buckling instability.
}  
\label{fig:compression-buckling}
\end{figure*}

\begin{figure*}[t!]
\centering
\includegraphics[scale=0.65]{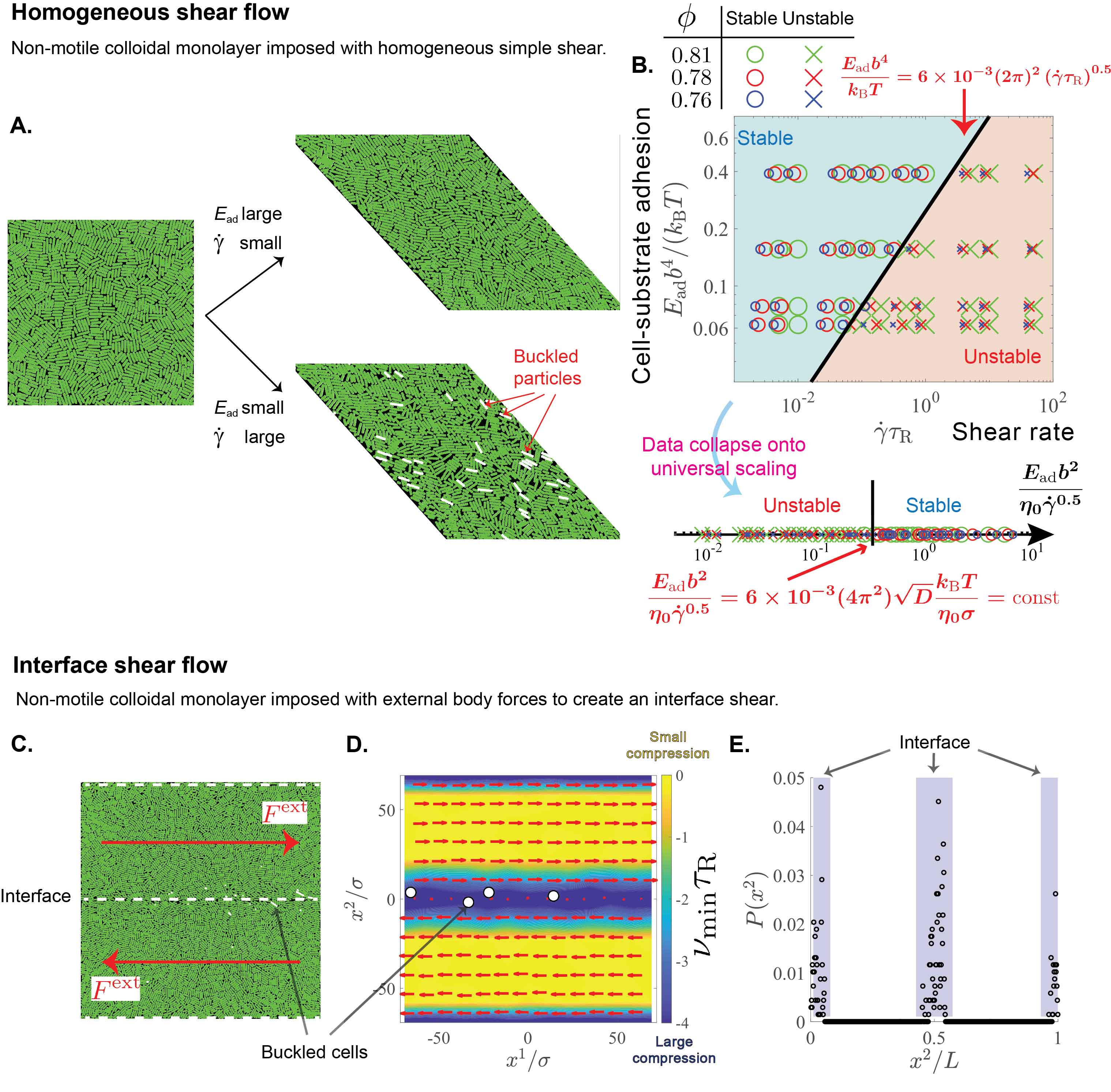}
\caption{
Molecular simulations validate the stability criterion derived from analytical theory for the case of incompressible shear flow; (A)-(B): Homogeneous shear flow, and (C)-(E): Interface shear flow.
(A) Snapshots of the simulations of non-motile, polydisperse spherocylinders imposed with homogeneous shear flow.  For large cell-substrate adhesion $E_{\mathrm{ad}}$ and small shear rates $\dot{\gamma}$, the particles remain stable within the monolayer.  For small $E_{\mathrm{ad}}$ and large $\dot{\gamma}$, we observe particles buckling out of plane (denoted by white colors).
(B) Results of stability across varying cell-substrate adhesion and imposed homogeneous shear rate, where circles and crosses indicate stable and unstable monolayers, respectively.  
The solid black line shows the predicted theory that divides the stable versus unstable regions, in excellent agreement with the MD simulations.  
Below, we collapse all data onto a universal scaling relation as predicted by the theory, thus demonstrating that shear stresses $\pi^{\alpha \beta}$ arising from in-plane incompressible shear flows lead to buckling of bacteria into the third dimension.
Buckling of monolayers occurs in regions of localized shear flows. 
(C) Snapshot of MD simulations of non-motile, polydisperse spherocylinders with an external body force. 
Particles on the upper half ($x^2 >0$) and lower half ($x^2 <0$) of the simulation box experience an equal and opposite body force $F^{\text{ext}}$ and $-F^{\text{ext}}$, respectively, creating two interfaces at $x^2 =0$ and $x^2  = \pm L_{x^2}/2$, where $L_{x^2}$ is the length of the simulation box.
(D) Snapshot of the flow fields (red arrows) at a given simulation timestep. 
Eigenvalues of the rate-of-strain tensor, $\mathbf{E} = (\nabla \mathbf{v} + (\nabla \mathbf{v})^T)/2$, give the magnitudes of compression and/or extension at each fluid element, and the smaller of the two eigenvalues, $\nu_{\text{min}}$, is plotted in color. 
Due to the nature of the body forces, large velocity gradients exist at the interfaces, creating localized shear stresses and thereby compression along one of the eigenvectors. 
White circular markers indicate particles that have buckled at this time step. 
(E) Probability of buckling along the $x^2$ direction.  
Peaks in the probability distribution (shaded) indicate that buckling occurs at interfaces of localized shear stresses, again demonstrating that shear stresses $\pi^{\alpha \beta}$ lead to nucleation into the third dimension.
}
\label{fig:molecular-testing}
\end{figure*}

\begin{figure*}[t!]
\centering
\includegraphics[scale=0.6]{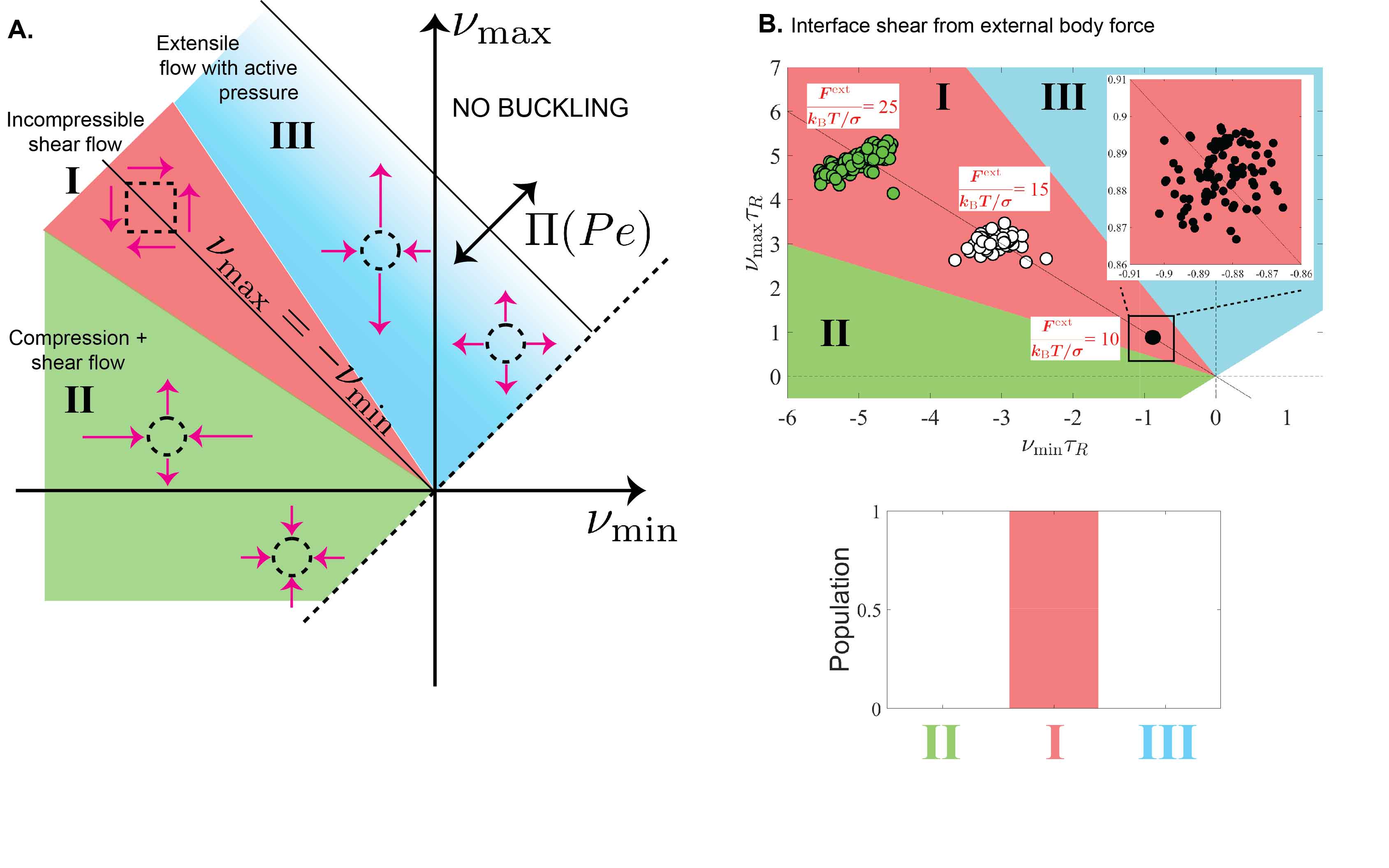}
\caption{
An eigenvalue-based metric to characterize the flow fields and identify the mechanisms behind buckling events. 
(A) Eigenvalue portrait showing different characteristic in-plane flows that can trigger buckling corresponding to distinct zones including incompressible shear flow (zone I), compression and shear flow (zone II), and extensile flows (zone III). 
Any region where $\nu_\text{max} + \nu_\text{min} \ne 0$ is a compressible flow.
(B) Top: Eigenvalue scatter plot computed from MD simulations of localized interface shear flows in Figs.~\ref{fig:molecular-testing}C, \ref{fig:molecular-testing}D and \ref{fig:molecular-testing}E for different values of external body forces (at a fixed particle-substrate adhesion), explicitly demonstrating that buckling is dominated by shearing flows (and therefore shear stresses) localized at the interface. 
Inset is a magnified view of the data corresponding to the smallest external force. 
Bottom: Histogram showing the population distribution of the buckled particles corresponding to the three zones contained in the eigenvalue-portrait (A), again showing quantitatively buckling mechanisms induced from incompressible shear flow.
}
\label{fig:eigenvalue-diagram}
\end{figure*}

\begin{figure*}[t!]
\centering
\includegraphics[scale=0.62]{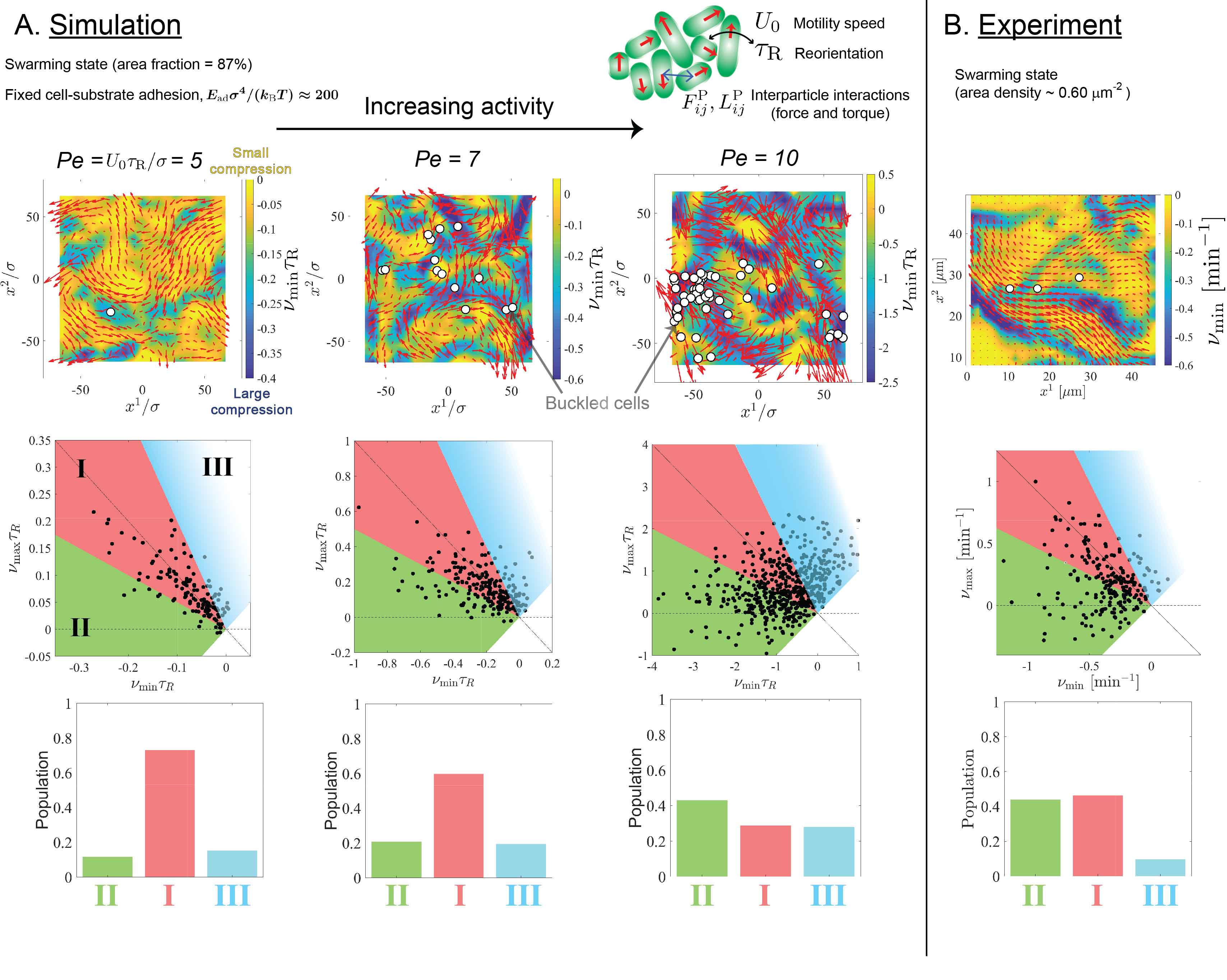}
\caption{Motility-induced, rate-dependent buckling of swarming monolayers in simulations and experiments. 
(A)  MD simulations of motile, polydisperse spherocylinders in the swarming state, below the onset pressure for glassy dynamics, across three different activities ($Pe$) for fixed cell-substrate adhesion.  
Top row: Snapshots of the flow fields (red arrows) and $\nu_{\text{min}}$ (colormap) within a region of the simulation.  The white circular markers indicate particles that have buckled at this time step. 
Middle row: Eigenvalue scatter plot corresponding to buckling. Low $Pe$ shows that majority of buckling events occur by motility-induced incrompressible shear flows. 
Increasing $Pe$ shows that buckling occurs from a broad variety of mechanisms, including inward and outward incompressible squeeze flows, and radial compressional or extensile flows.  
Buckling in the blue shaded region (extensile flows with $\nu_{\text{min}}+\nu_{\text{max}} > 0$) indicates that motility-driven in-plane pressure $\Pi$ can generate instability by overcoming stabilizing flows.
Bottom row: Histogram of different flow zones that resulted in buckling, color-coded to match the panels in the middle row, again showing buckling occurs from shear stresses at low $Pe$ to mechanisms mediated by different characteristic in-plane flows at high $Pe$. 
(B) Experiments of bacterial colonies in the swarming state. Top panel is a snapshot of flow fields (red arrows) and $\nu_{\text{min}}$ (colormap) within a region of the colony for cell density of $\sim 0.60 \mu \text{m}^{-2}$. 
Middle panel is the eigenvalue scatter plot across all cell densities tested in the swarming state (to the left of the glassy onset).  
Bottom panel shows the histogram of buckling events corresponding to the different flow zones. 
The similarity between experiments and active spherocylinder simulations at $Pe=10$ show that buckling in swarming \textit{P. aeruginosa} colonies are mediated by a broad variety of flow induced stresses and active pressures.
}
\label{fig:exps-sims}
\end{figure*}

For spherical active particles used for glass simulations in Fig.~\ref{fig:expts-glass} and Fig.~\ref{fig:DF theory}, we developed a custom code using hard disks, where the potential is singular and non-zero only at contact. 
To avoid interference by crystallization, we used a binary mixture of small and large particles with diameters $\sigma$ and $\sigma_1 = 1.4 \sigma $, and mole fractions $\chi_\sigma = 2/3$ and $\chi_{\sigma_1} = 1/3$, respectively. 
Dynamic structure factor was monitored to ensure that the relaxation time does not change with time.

For spherocylinder simulations, we used a GPU-enabled open source HOOMD-blue molecular dynamics package\cite{Anderson08,Glaser15}, where the spherocylinder particles use the rigid constraint functionality\cite{Nguyen11}.
Spherocylinders were created by a rigid assembly of overlapping spherical particles along their line of centers.
All pair interactions were modeled with a Weeks-Chandler-Andersen (WCA) potential \cite{Weeks71}, in which a Lennard-Jones potential is shifted upwards, truncated at the potential minimum of $2^{1/6} \sigma$ (such that the potential is purely repulsive), and assigned a well depth of $\epsilon$.
The spherocylinders have width $\sigma$ and mean side length $\ell_0 = 3\sigma$ (corresponding to the center-to-center distance of the first and last particles of the spherocylinder). 
Polydispersity was implemented in our simulations by normally distributing the particle length between seven discrete sizes.
Bacterial division is not modeled in this work in regards to the phenomena of motility induced buckling phenomena or glassy phenomena because of the large separation of timescales between motility ($\sim 20-35 \mu \text{m/min}$) and division rate ($\sim 1-2 \ \text{divisions/hr}$).
In all simulations, time steps were varied from $\Delta t= 10^{-6} - 10^{-5} (\sigma^2 / D)$, where $D$ is the Stokes-Einstein-Sutherland diffusivity based on the particle width $\sigma$.

\acknow{
We acknowledge Muhammad Hasyim for a critical reading of the manuscript. 
S.C.T. acknowledges support from the Miller Institute for Basic Research in Science at U.C. Berkeley.
K.K.M. acknowledges the support of University of California Berkeley.
K.K.M and S.C.T are also supported by Director, Office of Science, Office of Basic Energy Sciences, of the U.S. Department of Energy under contract No. DEAC02-05CH11231. 
We thank the Fleiszig Lab for generous donation of \textit{P. aeruginosa} strain PAO1.
}

\showacknow


\end{document}